\documentclass[aps,prl,a4paper,twocolumn,superscriptaddress,floatfix,showpacs]{revtex4}

\usepackage{graphicx}
\usepackage{dcolumn}
\usepackage{amsmath}
\usepackage{bm}
\usepackage{color}

\begin{document}

\title{Weakly interacting Bose gas in the one-dimensional limit}
\author{P.~Kr\"uger}
 \affiliation{Midlands Ultracold Atom Research Centre (MUARC), School of Physics and Astronomy, The University of Nottingham, Nottingham NG7 2RD, United Kingdom}
\author{S.~Hofferberth}
  \affiliation{Harvard-MIT Center for Ultracold Atoms, Department of Physics, Harvard University, Cambridge, MA 02138, USA}
\author{I.~E.~Mazets}
 \affiliation{Vienna Center for Quantum Science and Technology, Atominstitut, TU-Wien, Stadionallee 2, 1020 Vienna, Austria}
 \affiliation{Ioffe Physico-Technical Institute, 194021 St. Petersburg, Russia}
\author{I.~Lesanovsky}
 \affiliation{Midlands Ultracold Atom Research Centre (MUARC), School of Physics and Astronomy, The University of Nottingham, Nottingham NG7 2RD, United Kingdom}
\author{J.~Schmiedmayer}
 \affiliation{Vienna Center for Quantum Science and Technology, Atominstitut, TU-Wien, Stadionallee 2, 1020 Vienna, Austria}

\date{\today}

\begin{abstract}

We prepare a chemically and thermally one-dimensional (1d) quantum degenerate Bose gas in a single microtrap. We introduce a new interferometric method to distinguish the quasicondensate fraction of the gas from the thermal cloud at finite temperature. We reach temperatures down to $kT\approx 0.5\hbar\omega_\perp$ (transverse oscillator eigenfrequency $\omega_\perp$) when collisional thermalization slows down as expected in 1d. At the lowest temperatures the transverse momentum distribution exhibits a residual dependence on the line density $n_\mathrm{1d}$, characteristic for 1d systems. For very low densities the approach to the transverse single particle ground state is linear in $n_\mathrm{1d}$.
\end{abstract}

\pacs{39.90.+d, 03.75.Be}

\maketitle

Low dimensional systems are more fragile and prone to quantum and thermal fluctuations than their three-dimensional counterparts, so that the physics is drastically changed \cite{Giamarchi2004}. This is relevant in many different areas, but real systems are embedded in three dimensions. Cold atomic gases are well suited to synthesize low dimensional systems of interacting particles and to explore the influence of the frozen dimensions. The dimensionality can be controlled by the external trapping potential, quantum degeneracy can be reached by established cooling techniques, and inter-particle interactions can be tuned by controlling the density or a Feshbach resonance \cite{Bloch2008}.

A number of experiments have been performed with two- (2d) and one-dimensional (1d) cold atom systems, observing phenomena like the Kosterlitz-Thouless transition in 2d \cite{Hadzibabic2006,Schweikhard2007,Clade2009} or the `fermionization' of bosons in the 1d Tonks- Girardeau gas \cite{Kinoshita2004,Paredes2004}. To date, deeply 1d systems have been studied mostly in optical lattices where many 1d systems (typically $\sim 100$) are prepared at once \cite{Kinoshita2004,Stoeferle2004}. These systems are then characterized by averages over all `tubes'. In contrast, creating only one or two 1d clouds allows direct measurements of density \cite{Esteve2006} and phase \cite{Imambekov2009,Manz2010} fluctuations, facilitates studies of dynamical effects \cite{Hofferberth2007} and allows to probe the full distribution function of quantum variables \cite{Hofferberth2008,Gritsev2006,Kitagawa2010}. A recent experiment has shown that the exact Yang-Yang theory predicts a density distribution of a single 1d gas that matches the data better than ideal gas and mean field models \cite{Amerongen2008}. Earlier experiments have been performed in elongated macroscopic magnetic traps, where the systems exhibited some characteristic 1d properties while still in the 1d-3d crossover (chemical potential $\mu$ and temperature $kT>\hbar\omega_\perp$). Most notably, phase fluctuating `quasicondensates' were observed \cite{Det01,Richard2003,Trebbia2006}.

In this work we experimentally study the low temperature behavior of quantum degenerate Bose gases in the one-dimensional limit ($\mu,kT<\hbar\omega_\perp$). We study the {\em transverse} expansion of an individual degenerate 1d Bose gas, enabling us to explore the decreasing influence of 3d properties down to very low temperatures and chemical potentials. When $kT$ is lowered, the population of excited transverse modes is gradually reduced and the quasicondensate fraction grows. We directly investigate this process by observing the interference of two clouds after transverse expansion. The interference allows us to distinguish quasicondensed and thermal fractions of the gas as conventional techniques fail in 1d. We derive the temperature of the gas from the width ratio of these fractions and reach $kT\approx 0.5\hbar\omega_\perp$ as the lowest temperature in our experiment. Reaching such low $T$ allows us to observe the exponential suppression of thermalization rates when the dimensionality is reduced from three to one. In the low temperature regime, we show that even for $\mu<\hbar\omega_\perp$, inter-particle interactions lead to an admixture of 3d character at \textit{any} finite density. Only in the limit of vanishing density, the ground state width expected for the noninteracting gas is recovered.

In our experiment we prepare 1d Bose gases of $^{87}$Rb in a microscopic magnetic trap formed at a distance of $d\sim 10-30\,\mu$m from the surface of an atom chip \cite{Folman2000,Folman2002} as described in detail in \cite{Wildermuth2004}. At small atom-surface distances, small longitudinal variations of the trapping potential arising from slightly non uniform currents in trapping wires modulate the atomic density of the quantum degenerate gas. It is essential that these modulations are smaller than the chemical potential of the cloud and allow us to study linear density $n_\mathrm{1d}$ dependent effects \cite{Krueger2007b}. We are able to explore the entire density range from the detection threshold of $n_\mathrm{1d}\sim 3\,\mu$m$^{-1}$, corresponding to an interaction energy per particle of the order of $0.03\,\hbar \omega_\perp$ only ($\omega_\perp=2\pi\times 4$\,kHz), to well into the 1d--3d crossover regime $\mu \gtrsim \hbar \omega_\perp$ ($n_{1d}\approx 250\mu$m$^{-1}$). The length of the continuous gas reaches $L\approx 1$\,mm, corresponding to an aspect ratio of $L/a_\perp> 5000$ ($a_\perp= 170\,$nm is the transverse ground state size), the Lieb-Liniger parameter ranges between $\gamma=0.0014-0.12$.

\begin{figure}
    \includegraphics[angle=0, width= \columnwidth]{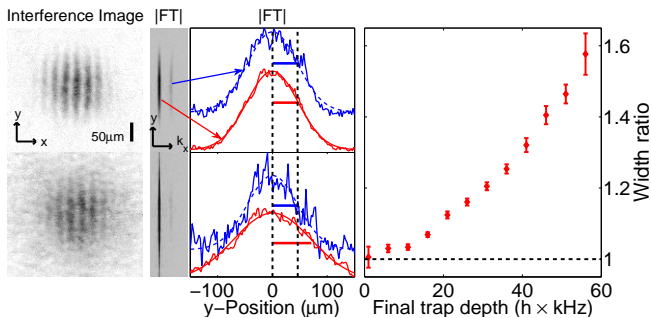}
    \caption{\label{fig:interference}
(color online) Interference of two 1d quasicondensates at relatively low (top left) and high (bottom left) temperatures. The absorption images were taken along the long direction ($z$) of the clouds after they had expanded from a double well trap in potential free time-of-flight. The absolute value of the Fourier transform (in arbitrary units) along the direction ($x$) of modulated density (fringes) is also depicted. The central panels show the density profiles at the interference fringe period (dashed blue line) and at zero spatial frequency (solid red line) and Gaussian fits (bars indicate the respective standard deviations $\sigma$). The panel on the right illustrates the increase in ratio between the widths of the cloud and the part of it that interferes as a function of final trap depth after the evaporative cooling process.}
\end{figure}

At finite temperature, an important question is which fraction of the gas is (quasi)condensed. The time-of-flight (TOF) technique of observation (in free expansion) of bimodal density profiles with a parabolic core and a surrounding thermal cloud has been very successful in studies of 3d BECs. In the case of 1d gases, however, this approach cannot be easily applied since both quasicondensate and thermal fraction expand transversely in a Gaussian fashion. Their widths after expansion hardly differ at sufficiently low $T$.

Our solution to this problem is to exploit the coherence property of a quasicondensate as revealed in an interference experiment. We trap a gas at sufficiently large distance from the surface ($d\sim 30\,\mu$m) to obtain completely smooth 1d clouds. We then split the gas using a radio frequency induced transformation of the trap into a double well as described in \cite{Schumm2005b,Lesanovsky2006,Hofferberth2006}. These two clouds are then released from the confining potential by suddenly switching off all trapping fields. After a free TOF expansion of 16 ms, the clouds overlap and form a high contrast interference pattern wherever the single particle ground state is macroscopically occupied.

Fig.\ \ref{fig:interference} shows two example images obtained in the way described above. One represents the low $T$ regime where essentially the entire cloud interferes, i.e.\ the sample is close to a pure quasicondensate. The other image clearly displays a core that interferes with high contrast, overlapped by a larger structureless thermal cloud. For quantitative analysis we perform a one-dimensional Fourier transform (FT) along each pixel row of each image. We determine the width of the coherent quasicondensate fraction by fitting a Gaussian to the first harmonic at $k_0$, the wave vector corresponding to the period of the fringe pattern. For comparison we fit a Gaussian to the overall density profile. Note that the conventional method of fitting bimodal profiles fails already at $k T \approx 3\hbar\omega_\perp$.

\begin{figure}
    \includegraphics[angle=0, width= \columnwidth]{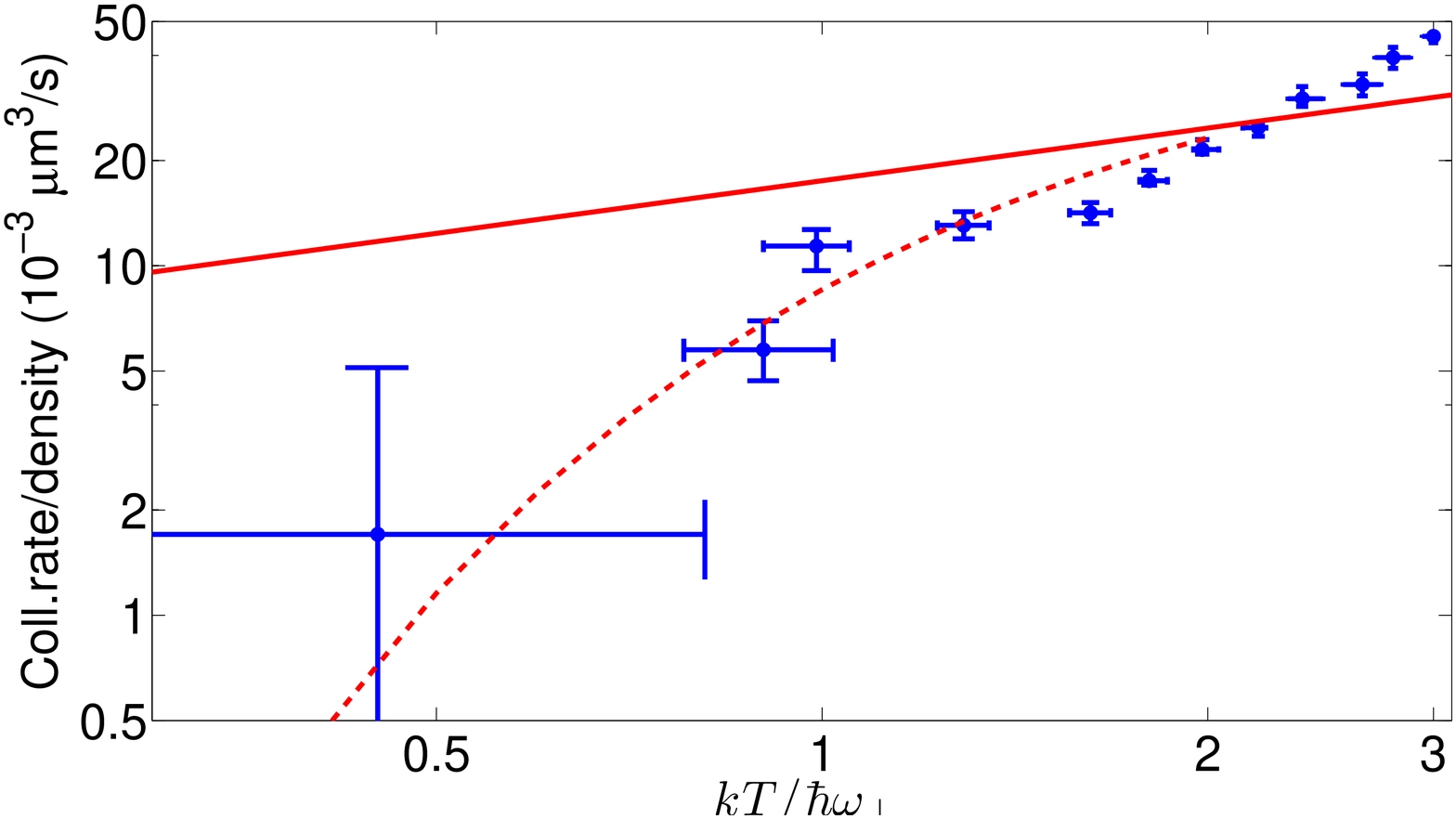}
    \caption{\label{fig:widths_if}
(color online) Collision rate devided by density as a function of temperature across the 1d-3d crossover. The lowest achieved temperatures ($kT\lesssim 0.5\hbar\omega_\perp$) appear to be limited by slow rethermalization due to inhibited elastic collisions. The increase in collision rate as temperature is increased does not follow the square root law (solid line shows best fit) that is valid in 3d, but is compatible with an exponential suppression as predicted in 1d (dashed line shows best fit for data within $kT < 2\hbar\omega_\perp$) \cite{Mazets2008}.}
\end{figure}

\begin{figure}
    \includegraphics[angle=0, width= \columnwidth]{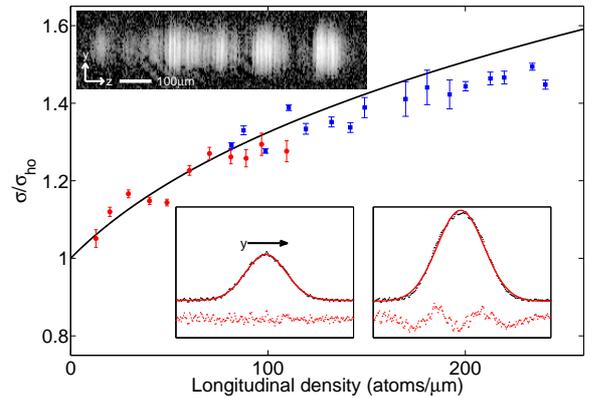}
    \caption{\label{fig:1d3dcrossover} (color online)
    Transverse width (Gaussian standard deviation, normalized to the single-particle ground state expanded size) of the expanded cloud after time-of-flight as a function of longitudinal line density of atoms throughout the quasi one-dimensional regime of $\mu\ll\hbar\omega_\perp$ and into the 1d-3d crossover. The points represent our data, measured in two different traps (blue squares and red circles, example data see upper inset). The line shows the theory without any adjustable parameters (Eq.\ \ref{eq:fabrice} \cite{Gerbier2004}). The lower insets illustrate that a Gaussian distribution models the data very well in the 1d regime (left, $n_\mathrm{1d}=100 \mu$m$^{-1}$), while systematic deviations become discernable in the 1d-3d crossover regime (right, $n_\mathrm{1d}=250 \mu$m$^{-1}$). Data, best Gaussian fits and residuals (magnified by a factor of 3) are shown.}
\end{figure}

The total transverse-momentum distribution of the expanded cloud is wider than its coherent part since the excited states of the radial motion contribute to the former, whereas the latter is determined by the properties of the ground state of the radial motion. The experimental parameter controlling the temperature of the gas is the final frequency $\nu_\mathrm{rf}$ of the radio frequency (rf) field (measured from the trap bottom) used to reduce the trap depth in the applied forced evaporation scheme. The ratio between thermal and coherent parts of the cloud as a function of the final depth of the trap $h\nu_\mathrm{rf}$ during the cooling process is plotted in Fig.\ \ref{fig:interference} (right). The coherent cloud's width indeed does not depend on $\nu_\mathrm{rf}$, small remaining variations (see also Fig.\ \ref{fig:1d3dcrossover}) can be attributed to variations in atom density as discussed below in more detail. At the lowest temperatures, the trap depth is set to very close to the chemical potential of the gas and partially depletes the quasicondensate.  At high temperature, the quasicondensate fraction decreases as the critical degeneracy temperature is approached. In contrast, the width of the thermal cloud increases with $h\nu_\mathrm{rf}$, as expected.

We assign a temperature $T$ to the gas according to the transverse ($y$) width $\sigma_y$ expected from an ideal Bose-Einstein gas. The temperature dependent width upon expansion was obtained by calculating the expectation value $\langle y^2\rangle$ using the grand canonical partition function of a 3d non-interacting Bose gas. Reflecting the experimental implementation, this gas is strongly harmonically confined in the two transverse dimensions. The chemical potential $\mu$ was fixed by the particle density at each $T$. We have verified that our experimental parameters are well in the dilute gas limit where interactions do not lead to detectable deviations from the ideal gas behavior \cite{Imambekov2007}. The lowest temperature we can reliably measure by this procedure is $kT \approx 0.5\hbar \omega_\perp$.

In evaporative cooling the ratio of $h\nu_\mathrm{rf}$ to the final temperature $kT$ (truncation parameter $\eta=h\nu_\mathrm{rf}/kT$) is a measure of the efficiency of the cooling. For $kT > \hbar \omega_\perp$ we find $\eta \sim 5$. As we approach colder temperatures $kT \sim \hbar \omega_\perp$,  $\eta$ drops significantly and reaches $\eta\approx 1$ for the lowest measured $T$. In our experiments the atoms were always released after a constant hold time $t=100\,$ms in the final trap.  This way $\eta$ can be used to quantify elastic collision rates $\gamma_c$ leading to thermalization of the gas \cite{Luiten1996}. Fig.\ \ref{fig:widths_if} displays $\gamma_c$ as a function of $T$. In 3d $\gamma_c$ is expected to be given by the constant elastic collisional cross section, the density $n$, and the mean particle velocity, i.e. $\gamma_c/n \propto \sqrt{T}$. The data clearly deviate from this scaling behavior. The faster drop of $\gamma_c$ for low $T$ can be related to reduced thermalizing binary collisions in the 1d regime. To illustrate the possibility of this effect being relevant in our case, we have fitted two model functions to our data: 1.) A square root behavior as expected in 3d; 2.) an exponential drop of thermalizing binary collision rates according to $\gamma_c/n\propto \exp{(-2\hbar\omega_\perp/kT)}$ \cite{Mazets2008} as expected in the low temperature regime ($kT\lesssim\hbar\omega_\perp$, the fit is for $kT<2\hbar\omega_\perp$). Three-body collisions start to be the dominant mechanism for $kT<0.3\hbar\omega_\perp$ \cite{Mazets2008,Mazets2010,Glazman2010}. We find that the strong reductions of thermalizing elastic binary collisions is consistent with our data. The associated reduction of thermalization rates \cite{Kinoshita2006} is likely to be the limiting factor in cooling further into the 1d regime.

We now investigate the low temperature limit $k_B T < \hbar\omega_\perp$ for a single 1d quasicondensate, in order to uncover the density dependent expanded cloud widths that cannot be seen at higher temperatures \cite{Goerlitz2001}. Here we exploit the density modulations that occur for small trap-surface distances ($d=10\,\mu$m). The inset of Figure \ref{fig:1d3dcrossover} shows a typical density profile of an expanding cloud after 6 ms TOF in a two-dimensional projection onto the $xz$ plane, so that the longitudinal ($z$) and one transverse ($y$) directions are resolved. The density profile is strongly modulated along $z$. The variations on longer length scales arise from a correspondingly inhomogeneous (and stable) potential. As the gas does not notably expand along the weakly confining $z$ direction during TOF, these features of the density profile are unaffected by TOF. We can confirm this by comparison to {\em in situ} images taken under the same experimental conditions. The known quasicondensate nature of the 1d quantum degenerate Bose gas causes random phase fluctuations in the trapped system that are converted to short range density fluctuations after TOF \cite{Det01,Richard2003,Imambekov2009,Manz2010}. This randomness allows us to eliminate this type of fluctuations by averaging over a few TOF images to ensure that the remaining modulation is due to actual density modulation in the trapped gas. In the transverse $y$ direction, on the other hand, the cloud's initial extension is comparable to the harmonic oscillator ground state size of $\sim 170\,$nm and cannot be resolved by the {\em in situ} imaging. After TOF, the size of the cloud along $y$ is then a direct measure of the energy stored in the trapped gas, since no longitudinal expansion occurs on the experimentally relevant timescales.

In the common case of a 3d harmonic trap when $\mu\gg\hbar\omega_\perp$, the interaction energy dominates the kinetic energy and a parabolic density profile emerges in the Thomas-Fermi approximation. In contrast, the zero point energy $\hbar\omega_\perp/2$ dominates in the 1d case when $\mu\ll\hbar\omega_\perp$, so that the shape of the distribution becomes Gaussian. By applying a local density approximation to averaged sets of images of the type shown in the inset of Fig.~\ref{fig:1d3dcrossover}, we obtain transverse TOF profiles of the cloud distribution as a function of $n_\mathrm{1d}$. We compare these distributions to best fits to Gaussian distributions (Fig.~\ref{fig:1d3dcrossover} insets) and confirm that as long as $n_\mathrm{1d}\lesssim 100\,\mu$m$^{-1}$, the Gaussian fits represent the data much more closely than the parabolic ones that would be expected in 3d. For larger densities $n_\mathrm{1d}\gtrsim 100\,\mu$m$^{-1}$, in the 1d-3d crossover, where $\mu\gtrsim\hbar\omega_\perp$, systematic deviations from the fitted model become discernable. Hence, Gaussian expansion is characteristic of the quasi 1d regime in which the shape of the momentum distribution of the single particle transverse ground state is directly observed.

Fig.~\ref{fig:1d3dcrossover} shows the width of the fitted Gaussian distributions as a function of $n_\mathrm{1d}$. Even down to the lowest linear densities, repulsive inter-particle interactions (scattering length $a_\mathrm{scat}=5.3\,$nm) broaden the distribution. Note that the scattering is still of 3d nature ($a_\mathrm{scat}\ll a_\perp=\sqrt{\hbar/m\omega_\perp}$), even in the regime where binary collisions cease to contribute to thermalization when $kT\ll\hbar\omega_\perp$.We compare our measured widths $\sigma(n_\mathrm{1d})$ to the theoretical predictions of the correction to the width $\sigma_0$ of the single-particle ground state case \cite{Salasnich2002,Gerbier2004}
\begin{equation}
    \frac{\sigma(n_\textrm{1d})}{\sigma_0} =
    \left(1+4a_{\mathrm{scat}}n_{\mathrm{1d}}\right)^\frac{1}{4}
    \label{eq:fabrice}
\end{equation}
and find excellent agreement in the absence of any fitting parameters for $\mu\lesssim\hbar\omega_\perp$. The data only starts to slightly deviate form this model when $\mu\approx\hbar\omega_\perp$, i.e.\,when the expansion of the gas is no longer purely Gaussian in the 1d--3d crossover regime. For large temperatures ($T \sim \hbar \omega_\perp$ or larger) the measured width saturates and stays at a constant value \cite{Goerlitz2001} given by the temperature, significantly above the curve shown in  Fig.~\ref{fig:1d3dcrossover}.

In conclusion, we have prepared quantum degenerate bosonic gases in the one dimensional limit. We use an interferometric technique to differentiate between quasicondensed and thermal part of the gas at finite temperatures. We find that the width of only the thermal cloud grows with temperature while the coherent part is temperature independent. Determining the temperature from the thermal cloud width and comparing it to the trap depth indicates a loss of cooling efficiency at very low temperatures $kT<\hbar\omega_\perp$, which can be attributed to suppressed thermalization in the 1d regime. In the low temperature regime ($kT\approx0.5\hbar\omega_\perp$), the transverse momentum distribution exhibits Gaussian shape. The width of this distribution, however, equals that of the single particle ground state only in the limit of vanishing line density when $\gamma\rightarrow\infty$. This implies that a true 1d gas is always a strongly correlated Tonks-Girardeau gas \cite{Kinoshita2004,Paredes2004}. Our measured widths follow the theoretical expectation for repulsive interactions.

We thank S. Wildermuth for help in the experiment. This work was supported by the European Union (SCALA), FWF, DFG, and EPSRC.

\end{document}